\documentstyle[]{article}

\begin{document}
\setlength{\unitlength}{0.17cm}
\title{Generalised coherent states for spinning relativistic particles}
\author{Charis Anastopoulos \thanks{anastop@physics.upatras.gr}, \\
\\Spinoza Instituut, Leuvenlaan 4, \\
3584HE Utrecht, The Netherlands \\ and \\
Department of Physics, University of Patras, \\
26194 Patras, Greece \\ (present address) } \maketitle

\begin{abstract}
We construct generalised coherent states of the massless and
massive representations of the Poincar\'e group. They are
parameterised by points on the classical state space of spinning
particles. Their properties are explored, with special emphasis on
the geometrical structures on the state space.
\end{abstract}
\renewcommand {\thesection}{\arabic{section}}
 \renewcommand {\theequation}{\thesection. \arabic{equation}}
\let \ssection = \section
\renewcommand{\section}{\setcounter{equation}{0} \ssection}

\section{Introduction}
The Poincar\'e group defines the basic symmetry of
non-gravitational physics. Every physical system on Minkowski
spacetime --quantum fields, in particular--  carries a
representation of the Poincar\'e group. Any such representation
may be written as a direct sum of irreducible representations. In
physical terms, an irreducible representation corresponds to an
elementary system characterised by group under study.

A full classification of the representations of the Poincar\'e
group was first achieved by Wigner, in his famous 1939 paper
\cite{Wig39}. Remarkably (but not unexpectedly), the irreducible
representations correspond to spinning particles. Excepting
unphysical and degenerate cases, the irreducible representations
either describe particles with finite mass $M$ and spin equal to
$\frac{n}{2} \hbar$, or massless particles of spin $\frac{n}{2}
\hbar$ and of either positive or negative helicity. This result
implies that any relativistic system, such as a quantum field, may
be analysed in terms of constituent particles, a fact making more
plausible the field-particle duality that lies at the heart of
quantum field theory.

The analysis of a relativistic system into elementary constituents
is not an exclusive quantum mechanical feature. It is also present
in classical mechanics. Any state space carrying a symplectic
Poincar\'e group action may be decomposed into elementary systems
(corresponding to transitive actions of the group) \cite{Sour}.
Similarly to the quantum case, these elementary systems correspond
to spinning particles. The only difference is that the quantum
description  forces the particle spin to take discrete values.

The classical state space $\Gamma$ and the quantum Hilbert space
$H$ of a physical system are related by means of the coherent
states, namely an overcomplete family of normalised vectors on
$H$, labelled by points of $\Gamma$ that satisfy a resolution of
the unity. The present paper deals with the construction of
generalised coherent states corresponding to the spinning
relativistic particles. For that purpose we exploit the fact that
the action of the unitary operators representing group elements on
a reference vector defines a set of generalised coherent states.
We make a convenient (Gaussian) choice for the reference vector
and show that the representations of the Poincar\'e group define
generalised coherent states for the spinning relativistic
particles, in full correspondence with the results of the
classical analysis. We then study the properties of those states.
A correspondence of classical functions to quantum operators needs
the existence of a resolution of the unity.  Even though the
Poincar\'e group leads to a fully covariant family of Hilbert
space vectors, a resolution of the unity  may be defined only by
restricting  on spatial hypersurfaces $\Sigma$. This procedure
breaks the full Poincar\'e covariance. This is the reason that the
natural position operators (like the Newton-Wigner one
\cite{NeWi49})  for relativistic particles do not transform
covariantly under the Poincar\'e group, even though the
corresponding classical functions do.

It needs to be emphasised that the massless and massive case are
very different. The state space for massless particles is not
simply the limit $M \rightarrow 0$ of the massive ones. It is a
different symplectic manifold, with different natural parameters
for the physical degrees of freedom, which may be conveniently
described in terms of naturally complex variables (twistors).

We place particular emphasis on the geometry of the classical
state space, which is induced by the generalised coherent states.
In particular, we identify a Riemannian metric on the (extended)
state space. Its role is twofold. First it determines the
resolution of phase space measurements thus  implementing the
Heisenberg uncertainty relation \cite{AnSav03}. Second, it is a
crucial ingredient of the coherent state path integral
\cite{KlDa84,Kla88}, because it defines a Wiener process through
which the path integral may be regularised.

This is not the first time that generalised coherent states of
relativistic particles have been constructed in the literature.
There exist, however, substantial differences between earlier work
and ours. We should emphasise that the generalised coherent states
we construct here are obtained from the representation theory of
the Poincar\'e group and the parameter state space is identified
with the classical symplectic manifold that described spinning
relativistic particles, and may be obtained, for instance, as
coadjoint orbits of the Poincar\'e group \cite{Sour}.

A complete and rigorous mathematical construction of a large class
of generalised coherent states of the Poincar\'e group has been
achieved in \cite{AGK96} - see also previous work \cite{AP86}.
Many families of generalised coherent states for massive
relativistic particles are constructed in these papers, without a
specification of the reference vector. The relevant parameter
space, however, is not the classical state space of a spinning
relativistic particle ${\bf R}^6 \times S^2$, but rather the state
space of a spinless relativistic particle ${\bf R}^6$, with the
spin degrees of freedom being treated as discrete variables. The
properties of those generalised coherent states are different from
the  ones of this paper (for example the distinction between
massless and massive particles).

Another construction of generalised coherent states of massive
spinning particles particles may be found in \cite{Or89}. This
work involves the representation theory of the group $SU(2)\times
SU(2)$ and they are therefore very different in structure from the
present ones. A construction of relativistic coherent states
within the general theory of wavelets may be found in reference
\cite{Kaiser}, in which the generalised coherent states are
labelled by points of a complexified Minkowski
spacetime--interpreted as the classical state space. Reference
\cite{CHV89} has dealt with the Moyal representation for spinning
relativistic particles, on the same state space with our
generalised coherent states. Finally, a precursor of our
construction for the massive spinless particles may be found in
\cite{AnSav03}.

The plan of the paper is as follows. In section 2 we provide the
necessary background. This involves the structure of the
Poincar\'e group, the basics of two-component spinors and some
basic facts about coherent states. In section 3 we construct the
generalised coherent states for massive particles and in section 4
for massless ones.

\section{Background}
\subsection{The Poincar\'e group}
The Poincar\'e group is the semi-direct product of the Lorentz
group and $R^4$, the Abelian group of spacetime translations on
Minkowski spacetime. An element of the Poincar\'e group is the
pair $(\Lambda^{\mu}{}_{\nu}, C^{\mu})$, which acts on points
$X^{\mu}$ of Minkowski spacetime as follows
\begin{eqnarray}
X^{\mu} \rightarrow \Lambda^{\mu}{}_{\nu} X^{\nu} + C^{\mu}.
\end{eqnarray}

In classical mechanics the state space is represented by a
symplectic manifold. For this reason we seek groups actions on
that manifold that preserve the symplectic structure . In quantum
mechanics the role of the state space is played by a complex
Hilbert space. We seek group actions that preserve the linearity
structure and the inner product of the Hilbert space, namely
 unitary group representations.

 When the Poincar\'e group
acts on the phase space $\Gamma$ of a physical system by
symplectic transformations, its Lie algebra is represented by
functions on  $\Gamma$ through the Poisson bracket. Writing the
generators of the Lorentz transformations as $M_{\mu \nu}$ and of
the spacetime translations  as $P^{\mu}$, we may define the
Pauli-Lubanski vector $W^{\mu}$.
\begin{equation}
W^{\mu} = \frac{1}{2} \epsilon^{\mu \nu \rho \sigma} P_{\nu}
M_{\rho \sigma}.
\end{equation}

Analogous operator relations hold in the quantum case.

The elementary systems -- the ones that correspond to transitive
actions classically and to unitary irreducible representations
quantum mechanically-- are classified by two physical quantities,
which are invariant under the action of the Poincar\'e group. The
first such invariant is the rest mass  $M := \sqrt{P_{\mu}
P^{\mu}}$ and the second is the spin $s := \sqrt{-\frac{1}{M^2}
W^{\mu} W_{\mu}}$. In the classical case spin takes any positive
value, while in quantum mechanics discrete values
 $s = \frac{n}{2} \hbar$, for any non-negative  integer $n$.

\subsection{Spinors}
In this section we  provide some basic expressions for the spinor
calculus, which are necessary in our treatment.

 The motivation for  spinors comes from the realisation
 that one may define a
self-adjoint complex $2 \times 2$ matrix $x_{A'A}$ for each
four-vector $X^{\mu}$ on Minkowski spacetime
\begin{equation}
X^{\mu} \rightarrow x_{A'A} = X^{\mu} (\sigma_{\mu})_{A'A}
\end{equation}
with $\sigma_0 = 1$ and $\sigma^i$ the Pauli matrices.

The inner product between two vectors reads
\begin{equation}
2 X^{\mu} Y^{\nu} \eta_{\mu \nu} = \epsilon^{AB}
\bar{\epsilon}^{A'B'} x_{A'A}y_{B'B},
\end{equation}

where$\epsilon = i \sigma_2$ is the totally antisymmetric tensor.

 From the above equation it follows that
\begin{equation}
\det x_{A'A} = X^{\mu} X_{\mu}
\end{equation}

For a null vector $X^{\mu}$, the determinant of the corresponding
matrix vanishes  and therefore
\begin{equation}
x_{A'A} =   \bar{c}_{A'} c_A,
\end{equation}
in terms of a non-zero element of ${\bf C}^2$, which is called a
{\em spinor}. Hence for each spinor $c_A$ there corresponds one
null vector
\begin{equation}
I^{\mu} = \bar{c} \sigma^{\mu} c,
\end{equation}
where the indices are suppressed and summation is implied.

If a spinor $c_A$ corresponds to a null vector $I^{\mu}$, so does
$e^{i \phi} c_{A}$. For this reason, the map from the space of
non-zero spinors  ${\bf C^2} - \{0\}$ to the space of null vectors
on Minkowski spacetime, is many-to-one. The map (2.7) then defines
a principal fiber bundle (the Hopf bundle), whose base space is
the space $V_+$ of future-pointing null vectors  (topologically $R
\times S^2$) with positive energy ($I^0
> 0$) \footnote{An analogous fiber bundle may be defined for null vectors with negative
energy.}, total space is ${\bf C}^2 -\{0\}$ (topologically $R
\times S^3$), fiber $U(1)$ and the projection map being defined by
means of equation (2.7).

If $I$ and $J$ are two null vectors with corresponding spinors $c$
and $d$ their product is
\begin{equation}
 2 I_{\mu}J^{\mu} = |c_A \epsilon^{AB} d_B|^2
\end{equation}

 In the following, we shall choose a reference cross-section of the Hopf bundle,
  by which a unique spinor
  $\iota$  represents
the  null vector $I^{\mu}$. The most convenient choice is to
consider spinors of the form $ \left( \begin{array}{c} e^{\rho} \\
e^{\rho} z \end{array} \right),$ for any real $\rho$ and complex
number $z$.

The Hopf bundle is non-trivial, hence this cross-section is not
global; it cannot be defined on the spinor $\left(
\begin{array}{c} 0 \\ 1 \end{array} \right)$. But for all other
spinors there exists
 an one-to-one map between future-directed null vectors and spinors, which reads explicitly.
\begin{eqnarray}
I^{\mu} \rightarrow \iota =
\left( \begin{array}{c} \sqrt{\frac{1}{2}(I^0+I^3)} \\
\frac{I^1 +i I^2}{\sqrt{2(I^0+I^3)}} \end{array} \right)
\end{eqnarray}
We can, nonetheless, make the definition of $\iota$ unique by
choosing $\iota = \left( \begin{array}{c} 0  \\ 1 \end{array}
\right)$ for $I^{\mu} = (1,-1,0,0)$.

On ${\bf C}^2$ there exists the defining action of the $SL(2,{\bf
C})$ group, i.e. of complex matrices with determinant one. For
each
 $\alpha \in SL(2,{\bf C})$ one may define an  element $\Lambda$ of the
Lorentz group
\begin{equation}
\Lambda^{\mu \nu} = \frac{1}{2} Tr(\alpha^{\dagger}\sigma^{(\mu}
\alpha \sigma^{\nu)} )
\end{equation}
The map is two-to-one since $\pm \alpha$ correspond to the same
Lorentz matrix $\Lambda$.

A pair of spinors $\iota, j$, such that $\iota^A \epsilon_{AB} j^B
= 1$ defines an {\em orthonormal null tetrad} of vectors
\begin{eqnarray}
I^{\mu} &=& \iota^* \sigma^{\mu}\iota \\
J^{\mu} &=& j^* \sigma^{\mu} j \\
m_1^{\mu} &=& \frac{1}{2} \left( \iota^* \sigma^{\mu}j +
j^* \sigma^{\mu}\iota \right) \\
m_2^{\mu} &=& \frac{1}{2i} \left( \iota^* \sigma^{\mu}j - j^*
\sigma^{\mu}\iota \right),
\end{eqnarray}
which satisfy the equations
\begin{eqnarray}
\eta^{\mu \nu} = \frac{1}{2} (I^{\mu}J^{\nu}+ I^{\nu}J^{\mu}) -
m_1^{\mu} m_1^{\nu} - m_2^{\mu} m_2^{\nu}.
\\
m_1^{\mu} m_2^{\nu} - m_1^{\nu} m_2^{\mu} = \frac{1}{2}
\epsilon^{\mu \nu \rho \sigma} I_{\rho} J_{\sigma}.
\end{eqnarray}

\subsection{Generalised coherent states}

 One may define generalised coherent states\footnote{In the present paper we consider
 as generalised coherent states any set of Hilbert space vectors
 labelled by points of a manifold, forming an overcomplete basis
 and possessing a resolution of the unity.}
  using the representation of a
group $G$ by unitary operators $\hat{U}(g), g \in G$ on a Hilbert
space $H$. Selecting a reference vector $|0 \rangle$ we
  may construct the vectors $\hat{U}(g)
|0 \rangle$. The usual choice for $| 0 \rangle$ is either the
minimum energy state or a vector that is invariant under the
maximal compact subgroup of $G$. We then define the
equivalence relation $~$ on $G$ as follows: \\ \\
$g \sim g'$ if there exists $e^{i \theta} \in
U(1)$ such that $\hat{U}(g) |0 \rangle = e^{i \theta} \hat{U}(g') |0 \rangle$.\\ \\
Defining the manifold $\Gamma = G/ \sim$, the map
\begin{equation}
[g] = z \in \Gamma \rightarrow \hat{U}(g) |0 \rangle \langle 0|
\hat{U}^{\dagger}(g)
\end{equation}
defines a set of generalised coherent states $|z \rangle$, which
possesses a resolution of the unity.

Through the generalised coherent states we may define a $U(1)$
connection on $\Gamma$
\begin{eqnarray}
iA = \langle z| dz \rangle,
\end{eqnarray}
which is familiar from the theory of geometric quantisation
\cite{Sour, Wood}. The closed
 two-form $\Omega = dA$ on
$\Gamma$ is in general degenerate, but  if it is not it equips
$\Gamma$ with the structure of a symplectic manifold. In that case
the Liouville form $\Omega \wedge \ldots \wedge \Omega$ defines an
integration measure on $\Gamma$ and suggests the existence of a
resolution of the unity.

The generalised coherent states also allow the introduction of a
Riemannian metric $ds^2$ on $\Gamma$
\begin{eqnarray}
ds^2 =  \langle d z|d z \rangle -  |\langle z|d z \rangle|^2.
\end{eqnarray}

 The  metric $ds^2$ defines a notion of distance on $\Gamma$ and incorporates the information
 about the uncertainty relation on phase space, namely the resolution
 in the determination of phase space properties. In previous work \cite{AnSav03}, we
proved that the condition $ \delta s^2 \sim 1$ is equivalent to
the Heisenberg uncertainty relations. The metric together with the
connection allow the determination of the coherent state
propagator $\langle z|e^{- i \hat{H}t}|z'\rangle$ by means of a
path integral

\begin{equation}
 \langle z''|e^{-i\hat{H}t}|z' \rangle =
 \lim_{\nu \rightarrow \infty} \int Dz(\cdot) e^{\nu t} e^{i \int A
 - i \int_0^t ds H  -
 \frac{1}{2 \nu} \int_0^t ds  g_{ij} \dot{z}^i \dot{z}^j },
\end{equation}
where the integral is over all paths $z(\cdot)$ such that $z(0) =
z'$ and $z(t) = z''$.

\section{Generalised coherent states for massive particles}
\subsection{The representation of the Poincar\'e group}
The unitary irreducible representations of the Poincar\'e group
may be constructed by Wigner's procedure.  We refer to the books
of Simms \cite{Simms} and Bogolubov et al \cite{Bog75} for a
comprehensive treatment, upon which we base the constructions of
the present paper.

The first step in Wigner's procedure involves the selection of a
reference unit timelike vector and identify its little group. We
choose the vector $n^{\mu} = (1,0,0,0)$. The corresponding little
group is the group $SO(3)$ of spatial rotations. Any element
$\Lambda$ of the Lorentz group may be written as a product
$\Lambda = \Lambda_I R$, where $R$ is a rotation --element of the
little group-- and $\Lambda_I$ is a boost taking $n^{\mu}$ to an
arbitrary unit timelike vector $I^{\mu}$
\begin{eqnarray}
(\Lambda_I)^{\mu}{}_{\nu} n^{\nu} = I^{\mu}
\end{eqnarray}

The boosts $\Lambda_I$ read explicitly
\begin{eqnarray}
(\Lambda_I )^{\mu}{}_{\nu} = \delta^{\mu}{}_{\nu} +
\frac{1}{I^0-1} (n^{\mu} - I^{\mu})(n_{\nu} - I_{\nu})
\end{eqnarray}

In the spinor representation $n^{\mu}$ corresponds to the {\em
unit} $2 \times 2$ matrix, while $\Lambda_I$ corresponds to the
hermitian matrix $\omega_I$
\begin{equation}
\omega_I = \sqrt{\tilde{I}} = \frac{1}{\sqrt{2(1 + I^0)}} ( 1 +
\tilde{I}),
\end{equation}
where $\tilde{I} = I^{\mu} \sigma_{\mu}$. The fact that $\omega_I$
is a positive matrix and the existence of a polar decomposition
for any matrix implies that an element of $SL(2, {\bf C})$ may be
written as $\omega_I u$, where $u$ a unitary $2 \times 2$ matrix.

The unitary irreducible representations of the $SL(2,{\bf C})$
group are classified by means of the unitary irreducible
representations of $SU(2)$, which is the universal cover of the
little group $SO(3)$. It is well known that the representations of
SU(2) are characterised by an integer $r$, which labels the
dimension of the representation's Hilbert space. We will denote by
$D^{(r)}(g)$ the unitary $r \times r$ matrix representing the
element $g \in SU(2)$.

To construct the representing Hilbert space we consider the space
$W_+$ of unit time-like vectors $\xi^{\mu}$ with positive value of
$\xi^0 = \sqrt{1 + {\bf \xi}^2}$, which is topologically
homeomorphic to ${\bf R}^3$. $W_+$ may be equipped with the
measure
\begin{equation}
d \mu_M(\xi) = M^2 d^4 \xi \delta(\xi^2 - 1) = M^2 \frac{d^3 {\bf
\xi}}{2 \xi^0},
\end{equation}
which are labelled by the value $M$ of the particle's rest mass.
The introduction of this measure  defines the Hilbert space ${\cal
L}^2(W_+,d \mu_M)$.

 The Poincar\'e group is represented on Hilbert spaces
$H_{M,r} = {\cal L}^2(W_+,d \mu_M) \otimes {\bf C}^r$, which
depend on the value of M and the integer $r$ labelling a
representation of $SU(2)$. The corresponding unitary operators
$\hat{U}(\alpha, X)$ are defined as
\begin{eqnarray}
[\hat{U}(\alpha, X)\Psi](\xi) = e^{- iM \xi \cdot X}
D^{(r)}(\omega^{-1}_{\xi} \alpha \omega_{\alpha^{-1} \cdot \xi})
\Psi(\alpha^{-1}\cdot \xi),
\end{eqnarray}
where $\alpha \in SL(2,{\bf C})$, $X^{\mu}$ correspond to the
Abelian group of spacetime translations, $\Psi(\xi) \in H_{M,n}$.
The expression $\alpha \cdot \xi$ denotes the adjoint action
$\alpha \tilde{\xi} \alpha^{\dagger}$ of $\alpha$ on the matrix
$\tilde{\xi}_{A'A}$ corresponding to the vector $\xi^{\mu}$.

\subsection{The construction}

We next select a reference vector to define  the generalised
coherent states. A vacuum state does not exist for free particle,
and also no vectors are invariant under the maximal compact
subgroup of the Poincar\'e group ($SO(3)$), unless the spin
vanishes. Hence, there exist no natural candidates for a reference
vector and  our choice will be guided by calculational
convenience. It should be noted that many of the results --such as
the structure of the symplectic manifold parameterising the
generalised coherent states-- do not depend on the explicit choice
of the reference vector. However the Riemannian metric on the
state space depends explicitly on that choice.

We choose a Gaussian vector $\psi_0 \in {\cal L}^2(W_+,d \mu_M)$,
\begin{equation}
\psi_0(\xi) = \frac{1}{M (\pi \sigma^2)^{3/4}} (2 n \cdot
\xi)^{1/2}  e^{- \frac{1}{2 \sigma^2} \xi \cdot {}^n\xi \cdot
\xi},
\end{equation}
where ${}^n\xi_{\mu \nu} = -\eta_{\mu \nu} + n_{\mu} n_{\nu}$.
This vector is centered around $\xi^i = 0 $ with a width equal to
$\sigma$.

We also choose a reference vector $| 0 \rangle_r$ on ${\bf C}^r$
\begin{eqnarray}
| 0 \rangle_r = \left(
\begin{array}{c} 1 \\ 0 \\. \\. \\. \\0 \end{array} \right)
\end{eqnarray}

 Then we may write a normalised reference vector on $H_{M,r}$
\begin{eqnarray}
\Psi_0(\xi) = \psi_0(\xi) \frac{D^{(r)}(\omega^{-1}_{\xi})| 0
\rangle_r }{\sqrt{{}_r \langle 0|\tilde{\xi}^{-1}| 0\rangle_r}} ,
\end{eqnarray}
where we extended the use of the symbol $D^{(r)}$ to refer to the
(non-unitary) representation of the $SL(2,{\bf C})$ associated
with the $r$-dimensional representation of $SU(2)$ \footnote{There
exist two possible extensions of $SU(2)$ representations to the
ones of $SL(2,{\bf C})$, depending on the embedding of $SU(2)$ in
$SL(2,{\bf C})$ in the fundamental representation. If $A$ is an
$SU(2)$ matrix one may define the map $A \in SU(2) \rightarrow A
\in SL(2,{\bf C})$, or the map $A \in SU(2) \rightarrow \epsilon
\bar{A} \epsilon^{-1}$, where $\epsilon = i \sigma_2$. The
reference vectors do depend that choice, however the properties of
the generalised coherent states are not affected. We shall employ
the first alternative in the present paper. } . The vector
$\Psi_0$ is centered around the momentum value $\xi^i = 0$, and
the spin pointing at the $(0,1,0,0)$ direction.

The action of $\hat{U}(\alpha, X)$ on $\Psi_0$ yields
\begin{eqnarray}
                 e^{- i M X \cdot \xi}  \psi_0(\alpha^{-1}\cdot \xi) \frac{D^{(r)}(\omega^{-1}_{\xi})
                   D^{(r)}(\alpha) | 0 \rangle_r}{\sqrt{{}_r \langle 0| D^{(r)}(\alpha)^{\dagger}
                   \tilde{\xi}^{-1} D^{(r)}(\alpha) |0
                   \rangle_r}}.
\end{eqnarray}

 If we effect the polar decomposition of the $SL(2,{\bf C})$
matrix $\alpha = \omega_I u$, the $SU(2)$ matrix $u$ will act on
the reference vector on ${\bf C}^r$ $|0 \rangle_r$, leading to the
generalised coherent states  of the group SU(2) $|{\bf \hat{m}}
\rangle_r$
\begin{equation}
D^{(r)}(u) |0 \rangle_r \rightarrow |{\bf \hat{m}} \rangle_r,
\end{equation}
 which are parameterised by a unit three-vector
${\bf \hat{m}}$ \cite{ACGT72}. If we denote by $\tilde{{\bf m}}$
the spinors corresponding to the three-vector ${\bf \hat{m}}$, the
inner product between the SU(2) generalised coherent states reads
\begin{equation}
{}_r\langle {\bf \hat{m}_1}|{\bf \hat{m}_2}\rangle_r = ({\bf
\tilde{m}^*_1 \cdot \tilde{m}_2})^r.
\end{equation}

In terms of the SU(2) coherent states, we define the following
family of Hilbert space vectors

\begin{eqnarray}
\Psi_{I,{\bf m},X}(\xi) = \frac{1}{M (\pi \sigma^2)^{3/4}} (2 I
\cdot \xi)^{1/2}  e^{- \frac{1}{2 \sigma^2} \xi \cdot {}^I\xi
\cdot
\xi - i M X \cdot \xi} \nonumber \\
\times \frac{D^{(r)}(\omega_{\xi}^{-1}) D^{(r)}(\omega_I) |{\bf
\hat{m}} \rangle_r}{\sqrt{{}_r \langle {\bf \hat{m}}|
D^{(r)}(\omega_I)^{\dagger}
                   \tilde{\xi}^{-1} D^{(r)}(\omega_I) |{\bf \hat{m}} \rangle_r}}.
\end{eqnarray}
The unit timelike four-vector $I^{\mu}$ is obtained by the action
of the Lorentz transformation corresponding to $\alpha$ on the
reference vector $n^{\mu}$. It represents the particle's
four-momentum normalised to unity. The unit three-vector
$\hat{m}^i$ corresponds to the direction of the particle spin on a
hypersurface normal to $n^{\mu}$.
 It is more convenient to employ the unit, spacelike, four-vector $J^{\mu}$ defined as
\begin{eqnarray}
J = \Lambda_I \left( \begin{array}{c} 0 \\ {\bf \hat{m}}
\end{array} \right)= \left(
\begin{array}{c}  {\bf \hat{m}} \cdot {\bf I}\\ ( \delta^{ij} - \frac{I^i I^j}{I^0-1})
\hat{m}^j
\end{array} \right).
\end{eqnarray}
 The four-vector $J_{\mu}$ satisfies $I \cdot J = 0$ and is related to the   Pauli-Lubanski
vector by $W^{\mu} = M \frac{r}{2} J^{\mu}$.

The family of vectors above may be represented by a ket $|X,I,J
\rangle_{M,r}$, which is parameterised by elements $(X,I,J)$ of
the nine-dimensional space $\Gamma_{M,r} = {\bf R}^7 \times S^2$.
The action of the Poincar\'e group leaves this set of Hilbert
space vectors invariant, in the sense that
\begin{eqnarray}
\hat{U}(\Lambda,0) |X,I,J \rangle = |\Lambda X, \Lambda I, \Lambda J \rangle \\
\hat{U}(0,Y) |X,I \rangle = |X+Y,I \rangle .
\end{eqnarray}
It should be emphasised that the spin degrees of freedom, encoded
in the normalised Pauli-Lubanski vector $J$ are continuous and
hence $|X,I,J \rangle_{M,r}$ is labelled by the parameters of the
classical state space, as appearing in the theory of
Konstant-Souriau.

  The space spanned by $X,I, J$ is odd-dimensional and for
this reason it is not expected to possess a resolution of the
unity. The vectors $|X,I,J \rangle_{M,r}$ do not define therefore
a family of generalised coherent states. One of the parameters in
the set of vectors above plays the role of time and it has to be
excised for a genuine family of generalised coherent states to be
constructed. Classically, one defines the space of true degrees of
freedom, by taking the quotient with respect to the action of the
subgroup of time translations -- the classical state space
$\Gamma_{M,r}$ then consists of all classical solutions to the
equations of motion, i.e. as the space of all orbits $(X,I,J)(s) =
(X_0 + M I_0 s, I_0, J_0)$, with $(X_0,I_0,J_0)$ a reference
point. We may then define a set of generalised coherent states
$|X(\cdot), I, J \rangle$, where $X(\cdot)$ is a path that solves
the classical equations of motion. With this parametrisation, the
set of generalised coherent states transforms covariantly --
similarly to equations (3.14--15)-- under the action of the
Poincar\'e group and it is equipped with a resolution of the
unity. To see this one may reduce the set of vectors $|X,I,J
\rangle$, by taking a fixed value of the parameter $t = n \cdot
X$, i.e. treating $t$ as an external parameter and not as an
argument of the generalised coherent states.

We  then define  the generalised coherent states at an instant of
time, i.e. a spacelike three-surface
  $\Sigma$, which is uniquely determined by the choices of $n^{\mu}$ and $t$.
  The generalised coherent states then
  depend on the spatial variables $x^i$ and $I^i$, which are the projections of
  $X$ and $I$ on $\Sigma$ together with the unit vector $\hat{m}^i$ of spin.
   These variables span
    the phase space of a single particle $T^*\Sigma \times S^2$,
    which is essentially the same with the covariant phase space
    spanned by the variables $X(\cdot), I, J$.

We  denote the generalised coherent states on $\Sigma$ as $|x, I,
\hat{m} \rangle_{\Sigma}$. The Poincar\'e group behaves as
follows: transformations that leave $\Sigma$ invariant (spatial
rotations and translations) preserve the generalised coherent
states, while the ones that take $\Sigma$ to  another surface
$\Sigma'$ (namely boosts and time translations) also take the set
of generalised coherent states into the  one associated to
$\Sigma'$.

We may explicitly compute
\begin{equation}
\int d^3 I d^3 x d^2\hat{m} \; \, \langle \xi | x, I, \hat{m}
\rangle_{\Sigma} {}_{\Sigma} \langle x, I, \hat{m}| \xi' \rangle =
\frac{1}{M^3} 2 \kappa \omega_{\xi} \delta^3(\xi - \xi'),
\end{equation}
a result that implies the existence of a resolution of the unity
\begin{equation}
 \kappa \hat{1} =  M^3 \int d^3I d^3x  d^2 \hat{m}
 |x, I, \hat{m} \rangle_{\Sigma} {}_{\Sigma} \langle x, I, \hat{m}| .
\end{equation}
Here $\kappa$ equals the mean value of energy in the vector
$\psi_0$
\begin{equation} \kappa = \int d \mu(\xi)
\omega_{\xi} |\psi_0|^2(\xi).
\end{equation}

Given a resolution of the unity, one may  provide natural
definitions of operators on $H_{M,r}$ in terms of functions on the
classical phase space. Hence for any function $f: T^* \Sigma
\times S^2 \rightarrow {\bf R}$, we may define the operator
$\hat{F}_{\Sigma}$ as
\begin{eqnarray}
\hat{F}_{\Sigma} = M^3 \int \frac{d^3I d^3x  d^2 \hat{m}}{\kappa}
f(x,I,\hat{m}) |x, I, \hat{m} \rangle_{\Sigma} {}_{\Sigma} \langle
x, I, \hat{m}|.
\end{eqnarray}
We should note here that the operators $\hat{F}_{\Sigma}$ do not
transform covariantly under the action of the Poincar\'e group. If
a Poincar\'e transformation takes a three-surface $\Sigma$ to a
three-surface $\Sigma'$, it {\em does not follow} that
$\hat{F}_{\Sigma}$ is related to $\hat{F}_{\Sigma'}$ by means of
the unitary operator corresponding to that Poincar\'e
transformation. In particular, if $\Sigma_t$ and $\Sigma_{t'}$ are
two surfaces, corresponding to two different moments of time with
respect to the same foliation, it does not follow that
\begin{eqnarray}
e^{i\hat{H}(t'-t)} \hat{F}_{\Sigma_t} e^{-i\hat{H}(t'-t)} =
\hat{F}_{\Sigma_{t'}}.
\end{eqnarray}

 For example, we may consider the position
operators $\hat{x}^i_{\Sigma}$, which represents length
measurements only on the surface $\Sigma$. The Hamiltonian
evolution yields
\begin{eqnarray}
e^{i\hat{H}(t'-t)} \hat{x}_{\Sigma_t} e^{-i\hat{H}(t'-t)} =
\hat{x}_{\Sigma_{t'}} + M \hat{I}_{\Sigma_{t'}} (t' - t)
\end{eqnarray}
It is often stated that the non-covariance of the position
operator implies that  particle position is not well-defined in
relativistic quantum mechanics. However, it needs to be noted that
the index $\Sigma$ does not refer to Heisenberg-time evolution,
but is a kinematical parameter determining the reference frame
that is involved in the specification of the corresponding
measurement. In the consistent histories approach to quantum
theory, the distinction between the kinematical and dynamical
aspect of the change in physical parameters has a nice
mathematical implementation \cite{Sav99a}, and there exists no
conflict with covariance in position being represented by means of
an one-parameter family of operators $\hat{x}^i_{\Sigma_t}$
\cite{Sav01a}.

We will next compare the method we followed here with that of
 reference
\cite{AGK96}, in which  spin is represented by discrete variables.
The starting point of \cite{AGK96} is the manifold $\Gamma_0$,
which is obtained as a quotient of the Poincar\'e group $G$ modulo
$SU(2) \times T$, where $T$ is the one-dimensional subgroup of
time translations. $\Gamma_0$ is essentially the classical phase
space of a massive, spinless relativistic particle (topologically
${\bf R}^6$). A fiber bundle $E(G,\Gamma_0, \pi)$  is then
naturally defined with total space the Poincar\'e group,
$\Gamma_0$ as base space and the projection $\pi$ defined through
the corresponding quotient. To construct a family of generalised
coherent states on $H_{M,r}$ one chooses $2r+1$ linearly
independent normalised vectors $| \eta^i \rangle$ on $H_{M,r}$ and
a section $\sigma$ of the bundle $E(G,\Gamma_0, \pi)$. The
generalised coherent states are then defined as
\begin{eqnarray}
|\xi, i \rangle = \hat{U}(\sigma(\xi))|\eta^i \rangle,
\end{eqnarray}
where $\xi \in \Gamma$. These  coherent states possess a
resolution of the unity. One may easily discern that the space
spanned by $|\xi, i \rangle$ is identical with $2s+1$ copies of
${\bf R}^6$: positions and momenta are continuously while  spin is
discrete.

The present method  considers the action of the full Poincar\'e
group on one reference vector of $H_{M,r}$.  The bundle
$E(G,\Gamma_0, \pi)$ is nowhere involved in this procedure either
explicitly or implicitly and for this reason our results do not
depend on the choice of a cross-section. The present method is the
standard one for obtaining generalised coherent states associated
to a group. We do  not assume here an  {\em a priori} distinction
between momenta-positions and spin degrees of freedom, and for
this reason spin and momentum are non-trivially intertwined in the
resulting parameter space. It is well known that this is the case
for spinning relativistic particles, For this reason it is very
difficult to relate directly the present construction with that of
 reference \cite{AGK96}, in which the spin degrees of
freedom are fundamentally distinguished from those of momentum.
The transformation properties under the Poincar\'e group are very
different.

 We should also remark that the coherent state parameter space in the present
method is not a quotient of the Poincar\'e group by any subgroup
(except for the trivial case $s=0$), but is defined by the
equivalence relation of vectors that correspond to the same ray
(see section 2.3). This parameter space can be identified with a
coadjoint orbit of the Poincar\'e group, which are classically
identified with the (unique) classical state space of massive
spinning particles.

\subsection{The coherent states' geometry}
\paragraph{Connection and symplectic form} We  now proceed to
study the geometry of the  parameter space for the generalised
coherent states. First, we evaluate the connection one-form. For
this purpose, it is more convenient to start with equation (3.9)
and parameterise the $SL(2,{\bf C})$ matrix $\alpha$ as
\begin{eqnarray}
\alpha = \left( \begin{array}{cc} a & b \\ c & e
\end{array} \right),
\end{eqnarray}
in terms of the complex numbers $a, b, c, e$, such that $ae - bc =
1$.

We then obtain
\begin{eqnarray}
d \Psi_{IJX}(\xi) = \left [\frac{\xi \cdot d I}{2 I \cdot \xi} -
\frac{\xi \cdot I \xi \cdot dI}{\sigma^2} - i M \xi \cdot dX
\right] \Psi_{IJX}(\xi) \nonumber \\ + \psi_{0}(\xi) \left(
\frac{D^{(r)}(\omega^{-1}_{\xi} d \alpha) | 0
\rangle}{\sqrt{\langle 0|D^{(r)}( \alpha^{\dagger}
\tilde{\xi}^{-1} \alpha) | 0 \rangle }}  \right. \nonumber \\
\left. - \frac{1}{2} \frac{\langle 0| D^{(r)}(\alpha^{\dagger}
\tilde{\xi}^{-1} d \alpha) | 0 \rangle + \langle 0| D^{(r)}(d
\alpha^{\dagger} \tilde{\xi}^{-1} \alpha) | 0 \rangle} {( \langle
0| D^{(r)}(\alpha^{\dagger} \tilde{\xi}^{-1}  \alpha) | 0 \rangle
)^{3/2}} D^{(r)}(\omega^{-1}_{\xi} d \alpha ) | 0 \rangle \right)
\end{eqnarray}

In order to compute the expression $\langle X,I,J|d| X,I,J
\rangle$, which involves integration over $d \mu_{M}(\xi)$ we
perform the change of variables $\tilde{\xi} \rightarrow
\alpha^{-1} \cdot \tilde{\xi}$. We also use the following relation
\begin{eqnarray}
{}_r\langle 0|D^{(r)}(\beta) | 0 \rangle_{r} = ({}_2 \langle 0 |
\beta| 0 \rangle_2)^r,
\end{eqnarray}
which enables us to compute all inner products in the fundamental
representation of $SU(2)$ on ${\bf C}^2$.

The first term in $\langle X,I,J|d| X,I,J \rangle$ reads
\begin{equation}
i M \kappa I^{\mu} d X_{\mu},
\end{equation}
while the second
\begin{eqnarray}
\frac{r}{2} [( e da - b dc)- (e^* d a - b^* dc^*)],
\end{eqnarray}
which may be written as
\begin{equation}
\frac{l}{2} [ \lambda_A \epsilon^{AB} d \mu_B - \lambda^*_{A'}
\epsilon^{A'B'} d \mu^*_{B'}],
\end{equation}
in terms of the two spinors
\begin{eqnarray}
\mu = \left( \begin{array}{cc} a & b \\ c & e \end{array}
\right) \left(\begin{array}{c}1 \\ 0 \end{array}  \right) =
\left( \begin{array}{c}a\\c \end{array} \right) \\
\lambda = \left( \begin{array}{cc} a & b \\ c & e \end{array}
\right) \left(\begin{array}{c}0 \\ 1 \end{array}  \right) = \left(
\begin{array}{c}b\\e \end{array} \right).
\end{eqnarray}
The spinor $\mu$ is obtained by a Lorentz transformation of the
spinor  $\left(\begin{array}{c}1 \\ 0 \end{array}  \right)$, which
corresponds to the null vector $(1,1,0,0)$. Hence, $\lambda$
corresponds to the null vector $I + J$. Similarly, the spinor
$\mu$ is obtained by a Lorentz transformation of the spinor
$\left(\begin{array}{c}0 \\ 1 \end{array}  \right)$, which
corresponds to the null vector $(1,-1,0,0)$. Hence, $\lambda$
corresponds to the null vector $I - J$. The two spinors satisfy
$\lambda_A \epsilon^{AB} \mu_B  = 1$. They, therefore, define a
null tetrad.

The final result is
\begin{eqnarray}
A = -  \kappa M I^{\mu} dX_{\mu} - \frac{i r}{2} [ \lambda_A
\epsilon^{AB} d \mu_B - \lambda^*_{A'} \epsilon^{A'B'} d
\mu^*_{B'}].
\end{eqnarray}
We may absorb $\kappa$ in a redefinition of the mass $M$ as $M'=
\kappa M$, or in a redefinition of the spacetime coordinates
$Y^{\mu} = \kappa X^{\mu}$. We shall prefer here the latter
alternative.

Under the gauge transformation $\mu \rightarrow e^{i \theta} \mu,
\lambda \rightarrow e^{-i \theta} \lambda$, the connection form
transforms as $A \rightarrow A + r d \theta$, while the two-form
\begin{eqnarray}
\Omega = M dI^{\mu} \wedge dY_{\mu} -i \frac{r}{2} [ d\lambda_A
\wedge \epsilon^{AB} d \mu_B - d \lambda^*_{A'} \wedge
\epsilon^{A'B'} d \mu^*_{B'}],
\end{eqnarray}
remains invariant. $\Omega$  may also be written in terms of the
vectors $I$ and $J$ as \cite{Sour}
\begin{equation}
\Omega = M dI^{\mu} \wedge dY_{\mu} - \frac{r}{4} \epsilon_{\mu
\nu \rho \sigma} I^{\mu}J^{\nu}(dI^{\rho} \wedge d I^{\sigma} - d
J^{\rho}\wedge dJ^{\sigma}).
\end{equation}

The two-form $\Omega$ is degenerate: the degenerate direction
corresponds to the vector field $ I^{\mu} \frac{\partial}{\partial
Y^{\mu}}$.

 Through the generalised coherent states, we have recovered the standard
form of the state space and symplectic structure of spinning
relativistic particles with non-zero mass.

\paragraph{The metric}  The
calculation  of the Riemannian metric on $\Gamma_{M,r}$ is
straightforward but tedious. The end result is the following
\begin{eqnarray}
ds^2 = ds^2_0 + \frac{i r}{4} \kappa M [ (\lambda \tilde{dX}
\mu^*) (\mu \epsilon d \mu) - (\mu \tilde{dX} \lambda^*)(\mu^*
\epsilon d \mu^*)] \nonumber \\ + \frac{r^2}{4} (1- v)|\mu
\epsilon d\mu|^2
\end{eqnarray}

Here $v$ denotes the constant
\begin{equation}
v = 2 \int d \mu_M(\xi) |\psi_0|^2(\xi) \frac{\xi^3}{\xi^0 +
\xi^3}.
\end{equation}
and $ds^2_0$ the corresponding metric for the spinless
relativistic particles
\begin{eqnarray}
ds^2_0 = - \frac{\omega}{3 \sigma^2} \eta_{\mu \nu}  dI^{\mu}
dI^{\nu} + K_{\mu \nu} dX^{\mu} dX^{\nu}.
\end{eqnarray}
The first term is the Riemannian metric on $W_+$  inherited from
the Lorentzian metric on Minkowski spacetime times a constant. The
parameter $\omega$ equals
\begin{equation}
\omega = \frac{1}{(\pi \sigma^2)^{1/2}}  \int_0^{\infty} \frac{d
\xi}{1 + \xi^2} e^{- \xi^2/ \sigma^2}
\end{equation}

The second term  involves the tensor
\begin{equation}
K^{\mu \nu} = \langle X,I|\hat{P}^{\mu} P^{\nu} |X,I \rangle -
\langle X,I|\hat{P}^{\mu} | X,I\rangle \langle X,I|\hat{P}^{\nu} |
X,I\rangle,
\end{equation}
which is the correlation tensor for the four-momentum on a
coherent state. Explicitly,
\begin{equation}
K_{\mu \nu} = M^2[(1 + \frac{2}{3}  \sigma^2 -
\kappa^2)I_{\mu}I_{\nu} -\frac{1}{6}  \sigma^2 \eta_{\mu \nu}].
\end{equation}

One may choose $\sigma^2 << 1$, in which case the reference vector
approaches weakly a delta function on momentum space. In that
case, the parameters $\kappa, \omega, v$ behave as
\begin{eqnarray}
\kappa = 1 + \frac{1}{4}  \sigma^2 - \frac{1}{16} \sigma^4 +
 O(\sigma^6) \\
 \omega = \frac{1}{2} + O(\sigma^2), \\
 v = O(\sigma^2).
\end{eqnarray}

This implies that the dominant contribution to the phase space
metric for small $\sigma^2$ is
\begin{eqnarray}
ds^2  = - \frac{1}{6 \sigma^2} \eta_{\mu \nu} dI^{\mu} dI^{\nu} +
M^2  \frac{\sigma^2}{6}  (I_{\mu} I_{\nu} - \eta_{\mu \nu} )
\delta X^{\mu} \delta X^{\nu} \nonumber \\
+ \frac{i r}{4} \kappa M [ (\lambda \tilde{dX} \mu^*) (\mu
\epsilon d \mu) - (\mu \tilde{dX} \lambda^*)(\mu^* \epsilon d
\mu^*)] + \frac{r^2}{4} (1- v)|\mu \epsilon d\mu|^2.
\end{eqnarray}

Note that this metric has a degenerate direction, which coincides
with that of the symplectic form (3.32).

In the particle's rest frame $I^i = 0 $ and for $t = 0$, the
spin-dependent terms in the metric read
\begin{eqnarray}
\frac{r}{2} M \kappa {\bf m} \cdot (d {\bf m} \times d {\bf x}) +
\frac{r^2}{4} d {\bf m} \cdot d {\bf m}.
\end{eqnarray}

The leading terms in the metric are quite important, as they are
less dependent on the details of the chosen reference vector. For
reasons of continuity, a small change in the reference vector
(with respect to the Hilbert space norm) will have a smaller
effect in the dominant terms. For this reason, the metric (3.43)
is the most suitable candidate for the path-integral calculation
of the coherent state overlap functional (2.20), which cannot be
analytically computed with our Gaussian wave functions.

It is well known that the knowledge of the overlap functional
enables one to fully reconstruct the information about the Hilbert
space and the coherent construction. Since we are using the metric
(3.43) and not the full metric (3.34) of the generalised coherent
states, the reference vector corresponding to that construction
will be different from the one we employed here. Still, the
geometric structure of the generalised coherent states will remain
the same.

\section{Generalised coherent states for massless particles}
\subsection{The representation of the Poincar\'e group}
The unitary irreducible representations of the Poincar\'e group
for zero mass are very different from the massive ones; they may
not be obtained as the $M \rightarrow 0$ limit of the massive
representations. For this reason the structure  of the
corresponding generalised coherent states are quite different.

We   follow again Wigner's procedure for the construction of the
group's representation. For that purpose, we select  a reference
null vector and identify its little group. It is convenient to
work in the spinor representation and take $\left(
\begin{array}{c}1 \\ 0 \end{array} \right) $ as a reference spinor.
The corresponding little group consists of all matrices $\left(
\begin{array}{cc} a & b \\ c & d \end{array}
\right)  \in SL(2, {\bf C})$ such that
\begin{eqnarray}
 \left( \begin{array}{cc} a & b \\ c & d \end{array} \right)
  \left( \begin{array}{c}1 \\ 0 \end{array} \right) =
  e^{i \phi} \left( \begin{array}{c}1 \\ 0 \end{array} \right),
\end{eqnarray}
for some phase $e^{i \phi}$. This is satisfied by all matrices of
the form
\begin{equation}
\left( \begin{array}{cc} e^{i \theta}&  e^{-i \theta}z \\ 0 & e^{-
i \theta} \end{array} \right)
\end{equation}
Each unitary representation of the little group defines uniquely a
unitary representation of the full Poincar\'e group.  The unitary
representations of this little group that are relevant to the
description of massless particles are one-dimensional and
correspond to the multiplication by a phase
\begin{equation}
\alpha= \left( \begin{array}{cc} e^{i \theta}& e^{-i \theta}z \\ 0
& e^{- i \theta} \end{array} \right) \rightarrow D_r(\alpha) e^{-
i r \theta},
\end{equation}
where $r$ is an integer that corresponds to the discrete values of
spin. The representations with opposite values of $r$ correspond
to particles with the same spin but opposite helicity.

Any element of $SL(2,{\bf C})$ may be written as a product of a
matrix of the form (4.3) with a matrix  of the form
\begin{equation}
\left( \begin{array}{cc} e^{\rho} & 0 \\ e^{\rho}z & e^{- \rho}
\end{array} \right)
\end{equation}

For each null vector $\xi^{\mu}$ we denote as $\omega_{\xi}$ the
unique matrix of type (4.4) that takes the reference spinor
$\left(
\begin{array}{c} 1 \\ 0 \end{array} \right)$ to the canonical spinor
$\tilde{\xi}$ associated to  $\xi$ \footnote{In this section we
denote as $\tilde{\xi}$ a spinor, while in the previous it denoted
the $2 \times 2$ matrix corresponding to $\xi^{\mu}$} .
In effect if $\tilde{\xi} = e^{\rho} \left( \begin{array}{c} 1 \\
z
\end{array} \right)$ then $\omega_{\xi} = \left( \begin{array}{cc}
e^{\rho} & 0 \\ e^{\rho}z & e^{- \rho} \end{array} \right).$

The massless representations are constructed on the Hilbert space
$H_0 = L^2(V_+, d \mu (\xi))$ of complex-valued, square-integrable
functions over the space $V_+$ of future-directed null vectors.
The measure $d \mu(\xi)$ is the unique Poincar\'e invariant
\begin{eqnarray}
d \mu(\xi) = \frac{d^3 \xi}{2 \xi},
\end{eqnarray}
where $\xi = \sqrt{{\bf \xi \cdot \xi}}$.

The representations are characterised by the integer $r$ of spin
\begin{equation}
[\hat{U}[\Lambda, X]  \Psi](\xi) = e^{i X \cdot \xi}
D_r[\omega_{\xi}^{-1} \alpha(\Lambda) \omega_{\Lambda^{-1}\xi} ]
\Psi(\Lambda^{-1} \xi),
\end{equation}
where $\alpha(\Lambda)$ is a $SL(2,{\bf C})$ matrix corresponding
to the Lorentz matrix $\Lambda$.

\subsection{The construction}
 We select a reference vector
sharply concentrated around a specific element of $V_+$,
conveniently chosen as $\xi^{\mu} = (1,0,0,1)$. We thus need to
identify smeared delta-functions on the space $V_+$.

Unlike the massive case,  $V_+$ has the topology ${\bf R} \times
S^2$, because the null vector $(0,0,0,0)$ is excluded. This
implies  that a (smeared) delta-function on $V_+$ factorises into
a product of a delta-function on ${\bf R}$ with a delta-function
on $S^2$. However, the identification of the component of
$\xi^{\mu}$ acting as coordinate on ${\bf R}$ and of the
components acting as coordinates on $S^2$ is not Lorentz
invariant. It depends on the choice of a reference timelike
vector. Choosing $n_R^{\mu} = (1,0,0,0)$, the coordinate $\xi =
n_{R}^{\mu} \xi_{\mu}$ takes values in $(0, \infty)$. Hence the
coordinate $\lambda = \log \xi^0$ runs across the full real line.

The sphere $S^2$ is essentially the "celestial sphere"
corresponding to the timelike direction $n_R$. The reference null
vector $(1,0,01)$ specifies a direction on this sphere
corresponding to the spacelike unit vector $m_R^{\mu} =
(0,0,0,1)$. The smeared delta function should be a function of
only the distance of the argument $\xi^{\mu}$ from the reference
vector $m_R^{\mu}$. It should be, therefore,  a function of
$m_R^{\mu} \xi_{\mu} = \xi \cos \theta$, where $\theta$ is the
angle between the three-vectors ${\bf \xi}$ and ${\bf m}_R$.

If we use as coordinates $\lambda$, $x = \cos \theta$ and $\phi$
(an azimuthal angle on the sphere running from $0$ to $2 \pi$),
the invariant measure becomes
\begin{equation}
d \mu(\xi) = e^{2 \lambda} d \lambda dx d \phi
\end{equation}

 It is convenient to employ a Gaussian
 as a smeared delta-function for the variable $\lambda$
\begin{equation}
f(\lambda) = \frac{1}{\sqrt{\pi \sigma^2} }
e^{-\frac{\lambda^2}{\sigma^2} - 2 \lambda}
\end{equation}

For the sphere $S^2$ recall that the delta-function with respect
to the north pole is given by
\begin{equation}
\delta(x) = \sum_{l = 0}^{\infty} \frac{2 l +1}{4 \pi} P_l (x)
\end{equation}
 where $x = cos \theta$ and $P_l$ the standard (unnormalised) Legendre polynomials.

A convenient choice for a smeared delta function
  is to truncate the series at some value $l = N$. So the smeared delta-function is
\begin{equation}
g(x) = \sum_{l=0}^N \frac{2 l + 1}{4 \pi} P_l(x).
\end{equation}
The benefit from this choice of smearing function is that for any
polynomial $f$ of $x$ of degree less or equal to N, we have
\begin{eqnarray}
2 \pi \int_{-1}^1 dx g(x) f(x) = f(1).
\end{eqnarray}

 With the previous choices of smeared delta functions
  we may write a reference vector on the Hilbert
space $H_0$
\begin{equation}
\Psi_0(\xi) = \sqrt{f}(\log n_R \cdot \xi) \sqrt{g}(\frac{m_R
\cdot \xi}{n_R \cdot \xi})
\end{equation}
 When the unitary operator $U[\alpha,X]$ acts on $\Psi_0$,  the argument of $\Psi_0$ goes from $\tilde{\xi}$ to $\alpha^{-1} \tilde{\xi}$.
  Since $\Psi_0$ is a function
  of $n_R\cdot \xi$ and $m_R \cdot \xi$, this transformation renders $\Psi_0$ into
  a function of $(\Lambda(\alpha) n_R)\cdot \xi$ and $(\Lambda(\alpha) m_R) \cdot \xi$, where $\Lambda(\alpha)$
  is the element of the Lorentz group corresponding to the $SL(2,{\bf C})$ matrix $\alpha$.
   The generalised coherent states
    depend on $n = \Lambda n_R$ and $m = \Lambda m_R$, which are unit timelike and spacelike
    vectors respectively that satisfy $n \cdot m = 0$. It is more convenient to employ a pair of  null vectors
     $I^{\mu} = n^{\mu} + m^{\mu}$,
$J^{\mu} = n^{\mu} - m^{\mu}$, which satisfy $I_{\mu}J^{\mu} = 2$.

The non-trivial part of the construction is the one referring to
the representation $D_r$ of the little group. If we write
$\xi^{\mu}$ in terms of its representative spinor $e^{\rho} \left(
\begin{array}{c} 1 \\ z \end{array} \right)$, and consider a
general $SL(2,{\bf C})$ matrix $\left( \begin{array}{cc} a & b \\
c & d \end{array} \right)$ we get
\begin{eqnarray}
D_r[\omega_{\xi}^{-1} \alpha  \omega_{\Lambda^{-1}\xi} ] = \left(
\frac{d - b z}{|d - bz|} \right)^{r}
\end{eqnarray}

The action of the $SL(2,{\bf C})$ matrix on $ \left(
\begin{array}{c} 0 \\ 1 \end{array} \right)$ gives $\left(
\begin{array}{c} b \\ d \end{array} \right)$. But $ \left(
\begin{array}{c} 0 \\ 1 \end{array} \right)$ corresponds to $n_R -
m_R^{\mu} = (1,0,0,-1)$ and hence $\left( \begin{array}{c} b \\ d
\end{array} \right)$ corresponds to $J^{\mu}$. Thus it can be
written as $j e^{i \chi}$ for some phase $\chi$. Taking this into
account we see that
\begin{equation}
D_r[\omega_{\xi}^{-1} \alpha  \omega_{\alpha^{-1}\xi} ] = \left(
\frac{\tilde{\xi}_A \epsilon^{AB} j^B e^{i \chi} }{|\tilde{\xi}_A
\epsilon^{AB} j^B|} \right)^{r}.
\end{equation}
However, the fact that $ad - bc = 1$ implies that $\chi$ must be
absorbed  in a redefinition of $j$ such that

\begin{equation}
\iota^A \epsilon_{AB} j^B = 1,
\end{equation}
so that the spinors $\iota$ and $j$ define a null tetrad. One
should note that --unlike the massive particles case -- the vector
$J^{\mu}$ is not here the normalised Pauli-Lubanski vector, since
the latter is a multiple of $I^{\mu}$ in the massless case.

 Eventually, using (4.14) we arrive at an expression for a set of
 vectors $| X,I,J \rangle_r $, from which we shall construct the
 generalised coherent states corresponding to the
 massless representations of the Poincar\'e
group
\begin{eqnarray}
\langle \xi| X,I,J \rangle_r = \Psi^{(r)}_{X,I,J} =  \left(
\frac{\tilde{\xi}_A \epsilon^{AB} j_B}{|\tilde{\xi}_A
\epsilon^{AB} j_B|} \right)^{r} e^{-i \xi \cdot X} \nonumber \\
\times \sqrt{f} \left( \log \frac{I+J}{2}\cdot \xi \right)
\sqrt{g}\left( \frac{-(I-J)\cdot \xi}{(I+J)\cdot \xi} \right)
\end{eqnarray}

The parameters $X,I,J$ of these vectors span a nine-dimensional
manifold, which we will call $\Gamma_{0,r}$. This is not, however,
the phase space of a classical system. We have to take into
account the fact that two different set of parameters correspond
to the same Hilbert space ray, i.e. that there might be a pair
$(X,I,J)$ and $(X',I',J')$ such that
\begin{equation}
\langle X,I,J|X',I',J' \rangle = e^{i \phi}
\end{equation}

Writing $X' = X+dX, I' = I + dI, J' = J+ dJ$, the above equation
reads
\begin{equation}
\langle X,I,J|d|X,I,J \rangle = i d \phi(X,I,J),
\end{equation}
or in terms of the U(1) connection $A$ of (2.18)
\begin{equation}
A - d \phi =0.
\end{equation}
One has, therefore, to excise all submanifolds of $M$ in which the
one-form $A$ becomes closed, or in other words remove all the
degenerate directions of the symplectic form $\Omega = dA$.

To compute $A$ we first write $d \Psi_{X,I,J}$
\begin{eqnarray}
d \Psi_{X,I,J} = - i \xi \cdot dX \Psi_{X,I,J}(\xi) \\
\nonumber
               + \frac{f'}{2f}(\log \frac{1}{2}\xi\cdot(I+J))
\frac{ \xi\cdot(dI+dJ)}{\xi \cdot(I+J)} \Psi_{X,I,J}(\xi)
\\ \nonumber
- \frac{g'}{g}( \frac{(I-J)\cdot \xi}{(I+J)\cdot \xi} ) \frac{(\xi
\cdot J)(\xi \cdot dI) - (\xi \cdot I) (\xi
\cdot dJ)}{[(I+J)\cdot \xi]^2} \Psi_{X,I,J}(\xi) \\
\nonumber + \frac{r}{2} \frac{(\tilde{\xi} \epsilon \tilde {J})^*
(\tilde{\xi} \epsilon dj) -(\tilde{\xi} \epsilon j)
(\tilde{\xi}\epsilon d j)^*}{|\tilde{\xi}\epsilon d j|^2 }
\Psi_{X,I,J}(\xi)
\end{eqnarray}

It is convenient to change variables to $\xi' = \Lambda^{-1} \xi$,
in order to compute the integral $ \int d\mu(\xi)
\Psi^*_{X,I,J}(\xi) d \Psi_{X,I,J}(\xi)$.  The reference null
directions become $I_R^{\mu} = (1,0,0,1)$ and $J_R^{\mu} =
(1,0,0,-1)$. In terms of these directions we can parameterise
$\xi'$ as
\begin{eqnarray}
\tilde{\xi}' = e^{\lambda} \left( \begin{array}{c} \sqrt{\frac{1+x}{2}} \\
\sqrt{\frac{1-x}{2}} e^{ i \phi} \end{array} \right),
\end{eqnarray}
where $x = \cos \theta$ refers to the angle between $\xi^i$ and
$m_R^i = (0,0,1)$.

The evaluation of the integral is now straightforward. The first
line of (4.20) gives a term $-i e^{\sigma^2/4} I^{\mu} dX_{\mu}$.
We can absorb the factor $e^{\sigma^2/4}$ into a redefinition of
$X^{\mu}$, i.e. write $Y^{\mu} = e^{\sigma^2/4} X^{\mu}$ so that
the first term reads $-i I^{\mu} dY_{\mu}$. The contributions of
the second and third term cancel each other, while the final term
contributes $\frac{r}{2} (\iota \epsilon dj - \iota^* \epsilon d
j^*)$. So the expression for the connection reads
\begin{equation}
A = - I^{\mu} dY_{\mu} - \frac{ir}{2} (\iota \epsilon dj - \iota^*
\epsilon d j^*)
\end{equation}

If we define the spinor
\begin{equation}
\omega_A = j_A + \frac{2i}{r} y_{A'A} \iota^{*A'},
\end{equation}
we obtain (up to a closed form)
\begin{equation}
A = \frac{ir}{2} (\iota^A \epsilon_{AB} d \omega^B - \iota^{*A'}
\epsilon_{A'B'} d \omega^{*B'}),
\end{equation}
giving the symplectic form
\begin{equation}
\Omega = \frac{ir}{2} (d \iota^A \wedge \epsilon_{AB}d \omega^B -
d \iota^{*A'} \wedge \epsilon_{A'B'}d \bar{\omega}^{A'})
\end{equation}

If we consider the spinor $\omega^A$ as a function of $Y$
--through equation (4.23)-- then it satisfies the {\em twistor
equation} (see for instance \cite{HuTo})
\begin{eqnarray}
\nabla_{A'}^{(A} \omega^{B)}(Y) = 0,
\end{eqnarray}
where $\nabla_{A'A} = \sigma^{\mu}_{A'A} \partial_{\mu}$. Note
that $\iota$ initially refers to the canonical expression (2.9)
for the spinor corresponding to the null vector $I^{\mu}$. Had it
been unrestricted, the pair $\iota^A, \omega_A$ would define an
element of the twistor space ${\bf T}$, namely the space of
solutions to equation (4.26).

 However, we may allow
variations of the phase of $\iota$. In particular, under the
transformation
\begin{eqnarray}
\omega_A &\rightarrow& \omega_A e^{i \theta} \nonumber \\
\iota^A &\rightarrow& \iota^A e^{-i \theta}
\end{eqnarray}
 the connection transforms
\begin{eqnarray}
A \rightarrow A - r d \theta,
\end{eqnarray}
which implies that the angle $\theta$ corresponds to a degenerate
direction of the symplectic two-form. Hence the generalised
coherent states'parameter space $\Gamma$ consists of equivalence
classes of pairs $(\iota^A, \omega_A)$ under the transformation
(4.27), which satisfy
\begin{equation}
\frac{1}{2}(\iota^{A} \epsilon_{AB} \omega^B + \iota^{*A'}
\epsilon_{A'B'} \omega^{*B'}) = 1.
\end{equation}
Equation (4.29) is due to  definition (4.23). In particular, this
equation implies that $\iota$ cannot vanish, in accordance with
the fact that $I^{\mu}$ may not take the value $(0,0,0,0)$.

  If we perform the transformation
\begin{eqnarray}
 \omega^A \rightarrow \zeta^A = \omega^A + \frac{2 i }{r} u j^A,
\end{eqnarray}
where $u = I^{\mu} Y_{\mu}$, we see that the spinor $\zeta^A$
satisfies $\iota^A \epsilon_{AB} \zeta^B = 1$. Hence the pair
$(\iota^A, \zeta^A)$ defines an orthonormal null tetrad. Moreover,
$\zeta^A$  transforms under (4.27) as
\begin{eqnarray}
\zeta^A \rightarrow \zeta^A e^{i \theta},
\end{eqnarray}
a fact that implies that the symmetry (4.27) of the symplectic
form corresponds to a rotation of the spacelike vectors $m_1$ and
$m_2$ of the null tetrad -- see equations (2.13-14). These vectors
 are not variables on the physical state space.  Consequently,
the space $\Gamma$ may be parametrised the null vectors $I^{\mu},
\zeta^{\mu}$ (with $I \cdot \zeta = 2$) corresponding to  the
spinors  $\iota^A, \zeta^A$, together with the parameter $u$.
Since $I^{\mu}$ cannot vanish, the topology of the resulting space
is ${\bf R}^4\times S^2$. Remarkably, the set of generalised
coherent states $|\iota,\zeta, u \rangle$ are parameterised by the
even dimensional symplectic manifold $\Gamma$, in a way that  does
not depend on the choice of a Lorentzian foliation. For this
reason, the generalised coherent states transform covariantly
under the action of the Poincar\'e group.
\begin{eqnarray}
\hat{U}(\alpha, 0) e^{i \chi} |\iota,\zeta , u \rangle \rightarrow
=
|\alpha \iota, \alpha \zeta, u \rangle \\
\hat{U}(1,C) |\iota,\zeta , u \rangle = e^{i \chi}| \iota, C\cdot
\zeta, u + I \cdot C \rangle,
\end{eqnarray}
where $C \cdot \zeta$ denotes the non-linear action of spacetime
translations on $\zeta$, by virtue of equations (4.23) and (4.30).
The phase $\chi$ depends on our phase convention about the
generalised coherent states. Clearly, the projection operators
$|\iota, \zeta, u \rangle \langle \iota, \zeta, u|$ transform in a
fully covariant manner under the Poincar\'e group.

\paragraph{ The phase space metric.} The determination of the phase space metric
involves extensive calculations. We here present the final result
\begin{eqnarray}
ds^2 = (e^{\sigma^2} - e^{\sigma^2/2}) (I \cdot dX)^2 \nonumber
\\+ \frac{1}{4} (1 + \frac{1}{2 \sigma^2} + 3 c_1) (I \cdot dJ)^2
- \frac{1}{2} (\frac{1}{4} c_2 +1) dI \cdot dI - \frac{1}{8} c_3
dJ \cdot dJ \nonumber \\+ (\frac{c_1}{4} -1) dI \cdot dJ  +
\frac{r^2}{2} F |j \epsilon
 d j|^2,
\end{eqnarray}
in terms of the coefficients
\begin{eqnarray}
c_1  &=& 2 \pi \int_{-1}^1 dx \frac{g'^2}{4g}(x) (1-x^2)^2 \\
c_2 &=& 2 \pi \int_{-1}^1 dx \frac{g'^2}{4g}(x) (1 - x^2) (1 +x)^2 \\
c_3 &=& 2 \pi \int_{-1}^1 dx \frac{g'^2}{4g}(x)  (1- x^2) (1 - x)^2 \\
F &=& 2 \pi \int_{-1}^1 dx g(x) \frac{1-x}{1+x}.
 \end{eqnarray}

 As the smearing parameters $\sigma^2 \rightarrow 0$ and $N \rightarrow 0$ the smearing
 function approaches weakly a delta function on momentum space. In
 that case the metric simplifies. However, the smeared delta function
 (4.10), which has been very convenient in our calculations, is of
 limited use in the explicit computation of the coefficients (4.35-38).
For this task we will employ a different smearing function on
$S^2$. This change does not affect the behaviour of the dominant
terms, except for the fact that they are written in terms of a
different smearing parameter. We, therefore, employ in equations
(4.35-38) the function
\begin{eqnarray}
g(x) = \frac{1}{2 \pi} C\frac{1+x}{(1-x)^2 + \epsilon^2}.
\end{eqnarray}
The coefficient is obtained from the normalisation condition $2
\pi \int_{-1}^1 dx g(x) =1$. Explicitly,
\begin{eqnarray}
C = [\frac{2}{\epsilon} \tan^{-1}\left( \frac{2}{\epsilon} \right)
+ \log \frac{\epsilon}{2}]^{-1}.
\end{eqnarray}

We may then evaluate the coefficients (4.35-38)
\begin{eqnarray}
c_1 &=& 4 + O(\epsilon) \\
c_2 &=& \frac{8}{\epsilon} + O(\epsilon^0) \\
c_3 &=& O(\epsilon) \\
F &=& \frac{\epsilon}{\pi} \log \frac{2}{\epsilon} + O(\epsilon^2)
\end{eqnarray}

Inspection of (4.34) shows that with the choice of $\epsilon = 8
\sigma^2$ the leading behaviour of the metric takes a rather
simple form
\begin{eqnarray}
ds^2 = \frac{\sigma^2}{2} (I \cdot dX)^2 + \frac{1}{8 \sigma^2}
(J_{\mu} J_{\nu} - \eta_{\mu \nu}) dI^{\mu} dI^{\nu} \nonumber \\
+ r^2 \frac{8 \sigma^2}{ \pi} \log \frac{1}{2 \sigma} |j \epsilon
 d j|^2
\end{eqnarray}

\section{Conclusions}

We constructed the generalised coherent states corresponding to
the physical unitary irreducible representations of the Poincar\'e
group. The space of parameters for these states correspond to the
classical symplectic manifold that describes spinning relativistic
particles. The description of these state spaces in terms of
generalised coherent states is perhaps more accessible (if less
elegant) to the particle physicist, because the standard classical
derivation involves rather advanced techniques of symplectic
geometry.

There are some differences and additions in our work, as compared
to the results that have appeared in the bibliography. We briefly
summarise  them here.
 \\ \\
${\bullet}$ Our generalised coherent states are obtained in a
straightforward manner from the  group representations of the
Poincar\'e group. The same procedure is followed, therefore, for
both the massless
and massive representations of the Poincar\'e group. \\
${\bullet}$ The parameter space of the resulting generalised
coherent states is identified with the classical state space of
spinning relativistic particles, which correspond to the coadjoint
orbits of the Poincar\'e group. This procedure highlights the
distinction
 between massive
and massless particles. \\
${\bullet}$ Our  choice for the reference vector allows us to
perform explicit calculations, such as the Riemannian matric on
state space, which is an essential ingredient of the
coherent-state path integral.

Our results imply that one may write a phase space representation
of quantum theory for spinning particles and for the fields
constructed from such particles. Geometric objects - such as the
$U(1)$ connection and the Riemannian metric on phase space will
play an important role in that description. It will be of great
technical and conceptual interest \cite{Ana03} to explore the
properties of quantum field theory in that particle
representation. The present paper provides a stepping stone in
that direction.

\section*{Acknowledgments}
The research was partly supported by a Marie Curie Fellowship of
the European Union.

\end{document}